\documentclass[prb,preprint,groupedaddress,showpacs]{revtex4}
\usepackage{graphicx} 
\begin{document} 
\bibliographystyle{prbrev}
 
\title{Spin-orbit effects in GaAs quantum wells
Spin-orbit effects in GaAs quantum wells:  Interplay
between Rashba, Dresselhaus, and Zeeman interactions
}

\author{Enrico Lipparini} 
\altaffiliation{Permanent address:
Dipartimento di Fisica, Universit\`a di Trento, and INFN,
38050 Povo, Trento, Italy}
\affiliation{Departament ECM, Facultat de F\'{\i}sica, 
and IN$^2$UB,
Universitat de Barcelona. Diagonal 647,
08028 Barcelona, Spain} 

\author{Manuel Barranco} 
\affiliation{Departament ECM, Facultat de F\'{\i}sica, 
and IN$^2$UB,
Universitat de Barcelona. Diagonal 647,
08028 Barcelona, Spain}

\author{Francesc Malet} 
\affiliation{Departament ECM, Facultat de F\'{\i}sica, 
and IN$^2$UB,
Universitat de Barcelona. 
Diagonal 647, 08028 Barcelona, Spain}

\author{Mart\'{\i} Pi} 
\affiliation{Departament ECM, Facultat de F\'{\i}sica, 
and IN$^2$UB,
Universitat de Barcelona. Diagonal 647,
08028 Barcelona, Spain}

\author{Lloren\c{c} Serra}
\affiliation{Departament de F\'{\i}sica, Universitat de les Illes Balears,
and Institut Mediterrani d'Estudis Avan\c{c}ats IMEDEA (CSIC-UIB),
 E-07122 Palma de Mallorca, Spain}

\date{\today}

\begin{abstract} 
The interplay between Rashba, Dresselhaus and Zeeman interactions 
in a quantum well submitted to an external magnetic field is
studied by means of an accurate analytical solution of
the Hamiltonian, including electron-electron interactions in a sum rule
approach. This solution allows to discuss
the influence of the spin-orbit coupling on some relevant
quantities that have been measured in inelastic light scattering
and electron-spin resonance experiments on  quantum wells.
In particular, we have evaluated the spin-orbit contribution to the spin splitting
of the Landau levels and to the splitting of charge- and spin-density
excitations. We also discuss how the spin-orbit effects change
if the applied magnetic field
is tilted with respect to the direction perpendicular
to the quantum well.
\end{abstract}


\maketitle

\section{Introduction}

The study of spin-orbit (SO) effects in semiconductor nanostructures has 
been the object of many experimental and 
theoretical investigations in the last few years, see e.g. 
Refs. \onlinecite{Can99,Ric99,And99,Vos00,Mal00,Rac97, 
Vos01,Fol01,Hal01,Ale01,Val02,Val202,Val302,Sch03,Kon05,Cal05} and Refs. 
therein. In spite of this,
 the extraction from measurements of the effective spin-orbit
coupling constant of both Dresselhaus\cite{Dre55} and 
Bychkov-Rashba\cite{Ras84,Pik95} SO interactions
is not a simple matter, since the SO corrections to the electron
energy spectrum in a magnetic field ($B$) are vanishingly small because
they correspond to second order effects in perturbation theory. 
Thus, few physical observables are sensitive enough to the
SO interactions and allow for
a quantitative estimate of their coupling constants.
One such observable is
the splitting of the cyclotron resonance (CR), which has been determined
in transmission experiments with far-infrared radiation,\cite{Man01}
and is due to the coupling between charge-density
and spin-density excitations.\cite{Ton04}
A less clear example is the  change in the Larmor frequency 
-spin splitting.\cite{Mal06} The spin splitting  has been observed
in electron-spin resonance\cite{Ste82,Dob88} and in inelastic
light scattering experiments.\cite{Dav97,Kan00}

In this work we extend our previous results\cite{Ton04,Mal06}
by obtaining an approximate, yet very accurate, analytical solution
of the quantum well SO Hamiltonian that contains both  Dresselhaus and
Bychkov-Rashba interactions. In the limit of 
high magnetic field, this solution
coincides with the results of second order perturbation theory,
 and allows to study the SO corrections to 
the Landau levels in a simple way, and to study the transitions
induced by an external electromagnetic
field acting upon the system.

This work is organized as follows. In Sec. II we present the general formalism
for the single-particle (sp) Hamiltonian.
These results are used in Sec. III to study the transitions 
caused by an external electromagnetic field.
The role of the electron-electron (e-e) interaction is discussed In Sec.
IV
within a sum rule approach. In Sec. V we discuss
the splitting of the Landau levels  and the appearance of charge- and
spin-density modes making, whenever possible, qualitative comparisons
with the experimental results.\cite{Man01,Dob88,Eri99,Sal01,Sih04}
A brief summary is presented in Sec. VI, and the generalization of
some of the expressions derived in Sec. II to the case of tilted
magnetic fields is presented in the Appendix.

\section{Single-particle states}

In the effective mass, dielectric constant approximation, the quantum well 
Hamiltonian $H$ can be written as $H=H_0
+ {e^2\over\epsilon}\sum_{i<j=1}^N{1\over\vert{\bf r}_i-{\bf r}_j\vert}$,
where $H_0$ is the one-body Hamiltonian  
consisting of the 
kinetic, Zeeman, Rashba and Dresselhaus 
terms
\begin{eqnarray} 
H_0 \equiv \sum_{j=1}^N [h_0]_j &=& \sum_{j=1}^N\left[\frac{P^+P^- + P^-P^+}{4m}+ 
\frac{1}{2}g^* \mu_B B \sigma_z \right. \nonumber\\ 
&+&
\left. \frac{\lambda_R}{2 i\hbar}(P^+\sigma_- - 
P^-\sigma_+)  
+ \frac{\lambda_D}{2\hbar}(P^+\sigma_+ + P^-\sigma_-)\right]_j
\; .
\label{eq1}
\end{eqnarray} 
$m=m^* m_e$ is 
the effective electron mass in units of the bare electron mass $m_e$, 
$P^{\pm}=P_x \pm i P_y$, 
$\sigma_{\pm}=\sigma_x \pm i\sigma_y$,
where the $\sigma$'s are the Pauli matrices, and
${\bf P}=-i\hbar\nabla+\frac{e}{c}{\bf A}$ represents the 
canonical momentum in terms of the vector potential ${\bf A}$, which in the
following we write in the Landau 
gauge, ${\bf A}= B (0,x,0)$, with ${\bf B}= \nabla \times{\bf A} = B \hat{\bf z}$.
The second term in Eq. (\ref{eq1}) is the Zeeman energy, where
$\mu_B = \hbar e/(2 m_e c)$ is the Bohr magneton, and $g^*$ is the effective 
gyromagnetic factor. The third and fourth terms are the usual Rashba and
Dresselhaus interactions, respectively. Note that 
for bulk GaAs, taken here as an example,
$g^*=-0.44$, $m^*=0.067$, and the dielectric constant 
is $\epsilon=12.4$.
To simplify the expressions, in the following we
shall use effective atomic units 
$\hbar=e^2/\epsilon=m=1$. 

Introducing the operators
\begin{equation}
\label{eq2}
a^{\pm}={1\over\sqrt{2\omega_c}}P^{\pm}
\end{equation}
with $[a^-,a^+]=1$ and $\omega_c=e B/c$ being the cyclotron frequency, the 
sp  Hamiltonian $h_0$ 
can be rewritten as
\begin{equation} 
h_0/\omega_c = \frac{1}{2}(a^+a^- + a^-a^+) - {1\over2}{\omega_L\over\omega_c}\sigma_z
-\frac{1}{2}\, i\tilde{\lambda}_R(a^+\sigma_- -
a^-\sigma_+) 
+ \frac{1}{2}\, \tilde{\lambda}_D(a^+\sigma_+ + a^-\sigma_-) \;\;\;\; ,
\label{eq3}
\end{equation} 
where $\omega_L=|g^* \mu_B B|$ is the Larmor frequency
and $\tilde{\lambda}_{R,D}=\lambda_{R,D}\sqrt{{2\over\omega_c}}$.
For the spinor $|\phi \rangle \equiv \left(\begin{array}{c}
\stackrel{\textstyle \phi_1}{\textstyle \phi_2} \end{array} \right)$
(we shall  use `1' for the top component and `2' for the bottom component of any 
spinor), the Schr\"odinger equation $h_0 |\phi\rangle =\varepsilon |\phi \rangle$ 
adopts the form
\begin{equation}
\left[ \begin{array}{cc}
\frac{1}{2}(a^+a^- + a^-a^+) - \omega_L/(2\omega_c)-\varepsilon &
i\tilde{\lambda}_R a^- + \tilde{\lambda}_D a^+ \\
-i\tilde{\lambda}_R a^+   +\tilde{\lambda}_D a^- & 
\frac{1}{2}(a^+a^- + a^-a^+) + \omega_L/(2\omega_c)-\varepsilon
 \end{array}\right]\left(\begin{array}{c}\phi_1\\\phi_2\end{array} \right)=0
\;\:\;\; .
\label{eq5}
\end{equation}
We expand $\phi_1$ and $\phi_2$ into oscillator states $\vert n\rangle$ 
as $\phi_1=\sum_{n=0}^\infty a_n\vert n\rangle$, 
$\phi_2=\sum_{n=0}^\infty b_n\vert n\rangle$ , on which $a^+$ and $a^-$ act in the 
usual way, i.e.,
 $\frac{1}{2}(a^+a^- + a^-a^+)\vert n\rangle=(n+\frac{1}{2})\vert n\rangle$,
$a^+\vert n\rangle=\sqrt{n+1}\vert n+1\rangle$, $a^-\vert n\rangle=\sqrt{n}\vert 
n-1\rangle$,  and $a^-\vert 0\rangle=0$. This yields the infinite system of equations
\begin{eqnarray}
(n+\alpha-\varepsilon)b_n
-i\tilde{\lambda}_R\sqrt{n}a_{n-1}+\tilde{\lambda}_D\sqrt{n+1}a_{n+1}=0
\nonumber\\
(n+\beta-\varepsilon)a_n
+i\tilde{\lambda}_R\sqrt{n+1}b_{n+1}+\tilde{\lambda}_D\sqrt{n}b_{n-1}=0  
\label{eq6}
\end{eqnarray}  
for $n\ge0$, with $a_{-1}=0$, $b_{-1}=0$, and
$\alpha=(1+\omega_L/\omega_c)/2$, $\beta=(1-\omega_L/\omega_c)/2$.

\subsection{Case in which either \mbox{\boldmath $\lambda_R =0$}, or
\mbox{\boldmath $\lambda_D =0$}.}

When only the Rashba or Dresselhaus terms are considered,
Eqs. (\ref{eq6}) can be exactly solved.\cite{Ras60,Das90,Fal93,Sch03} 
For the sake of completeness, we give here the corresponding results.
In the $\lambda_D=0$ case, combining Eqs. (\ref{eq6}) one obtains
\begin{eqnarray}
\left[(n+\alpha-\varepsilon)
(n-1+\beta-\varepsilon)-n\,{\tilde{\lambda}_R}^2\right]
\, b_n &=& 0
\nonumber\\
\left[(n+\alpha-\varepsilon)
(n-1+\beta-\varepsilon)-n\,\tilde{\lambda}_R^2
\right] a_{n-1} &=& 0 \;\;\;\; ,
\label{eq7}
\end{eqnarray}
either of which yields the energies
\begin{equation}
\label{eq71}   
\varepsilon^{\pm}_n=n\pm\sqrt{{1\over4}
\left(1+{\omega_L\over\omega_c}\right)^2
+{2\over\omega_c}\lambda^2_R \,n} \;\;\;\; .
\end{equation} 
One also obtains
\begin{equation}
\label{eq8}
(n-1+\beta-\varepsilon^{\pm}_n)\,a^{\varepsilon^{\pm}_n}_{n-1}
=-i\tilde{\lambda}_R\sqrt{n}\,b^{\varepsilon^{\pm}_n}_{n} \;\;\;\; ,
\end{equation} 
which together with the normalization condition 
$\vert a^{\varepsilon^{\pm}_n}_{n-1}\vert^2
+\vert b^{\varepsilon^{\pm}_n}_{n}\vert^2=1$ 
solves exactly the problem
[for $n=0$, $a_{-1}=0$, $b_0=1$, and
$\varepsilon_0=\frac{1}{2}(1+\omega_L/\omega_c)$].

Eqs. (\ref{eq7}) indicate that in the series expansion of the spinor
$|\phi \rangle$, only one $a_i$ and one $b_i$ coefficient appears.
Specifically, 
\begin{equation}
 |n_d \rangle =
 \left(\begin{array}{c}
a^{\varepsilon_n^+}_{n-1}\, |n-1 \rangle \\
b^{\varepsilon_n^+}_{n} |n\rangle
 \end{array} 
 \right) \;\; ; \;\;
 |n_u\rangle=
 \left(\begin{array}{c}
a^{\varepsilon_{n+1}^-}_n\, |n \rangle \\
b^{\varepsilon_{n+1}^-}_{n+1} |n+1 \rangle
 \end{array} 
 \right) \;\;\; .
\label{eq8c}
\end{equation}
In the limit of zero spin-orbit, the spinors
$|n_d \rangle$ and $|n_u \rangle$ become
$|n\rangle$$\left(\begin{array}{c} \stackrel{\textstyle 0}{1}
\end{array} \right)$ and 
$|n\rangle$$\left(\begin{array}{c} \stackrel{\textstyle 1}{0}
\end{array} \right)$, respectively.
The exact expressions for the $a_i$ and $b_i$ coefficients entering
Eq. (\ref{eq8c}) are easy to work out. Expressions valid up to
$\lambda_{R,D}^2$ order are given in the next subsection.

The $\lambda_R=0$ case can be worked out similarly. One obtains the
secular equation
\begin{equation}
\label{eq9}
(n+\beta-\varepsilon)(n-1+\alpha-\varepsilon)-n\,{\tilde{\lambda}_D}^2=0
\end{equation}
which yields
\begin{equation}
\label{eq91}
\varepsilon^{\pm}_n=n\pm\sqrt{{1\over4}
\left(1-{\omega_L\over\omega_c}\right)^2 +{2\over\omega_c}\lambda^2_D
\,n} \;\;\;\; .
\end{equation}
One also obtains
\begin{equation}
\label{eq10}
(n-1+\alpha-\varepsilon^{\pm}_n)\,b^{\varepsilon^{\pm}_n}_{n-1}
=-\tilde{\lambda}_D\sqrt{n}\;a^{\varepsilon^{\pm}_n}_{n} \;\;\;\; ,
\end{equation}
which together with the normalization condition
$\vert a^{\varepsilon^{\pm}_n}_{n}\vert^2
+\vert b^{\varepsilon^{\pm}_n}_{n-1}\vert^2=1$ solves exactly the
problem (in this case, for $n=0$, $b_{-1}=0$ and $a_0=1$).

Again, in the series expansion of the spinor
$|\phi \rangle$, only one $a_i$ and one $b_i$ coefficient appears:
\begin{equation}
 |n_d \rangle =
 \left(\begin{array}{c}
a^{\varepsilon_{n+1}^-}_{n+1}\, |n+1 \rangle \\
b^{\varepsilon_{n+1}^-}_{n} |n\rangle
 \end{array} 
 \right) \;\; ; \;\;
 |n_u \rangle =
 \left(\begin{array}{c}
a^{\varepsilon_n^+}_n\, |n \rangle \\
b^{\varepsilon_n^+}_{n-1} |n-1 \rangle
 \end{array} 
 \right) \;\;\; ,
\label{eq10c}
\end{equation}
and the same comments as before apply.

\subsection{General case when \mbox{\boldmath $\lambda_R \neq0$} and
\mbox{\boldmath $\lambda_D \neq 0$}.}

If both terms are simultaneously considered, the SO interaction
couples  the states of all Landau levels, and an exact analytical
solution to Eqs. (\ref{eq6}) is unknown, and likely does not exist. 
We are going to find an approximate solution that
in the  $\lambda^2_{R,D}/\omega_c\ll1$ limit coincides
with the results of second order perturbation theory,
i.e., it is valid up to $\tilde{\lambda}_{R,D}^2$ order, and it is
quite accurate as compared with exact results obtained numerically.
Combining Eqs. (\ref{eq6}), one can write
\begin{eqnarray}
\label{eq12}
\left[n+\alpha-\varepsilon-{\tilde{\lambda}_R}^2{n\over 
n-1+\beta-\varepsilon}-{\tilde{\lambda}_D}^2{n+1\over 
n+1+\beta-\varepsilon}\right]b_{n}=
\nonumber\\
\nonumber\\
-i\tilde{\lambda}_R\tilde{\lambda}_D\left[{\sqrt{n(n-1)}
\over n-1+\beta-\varepsilon}\,b_{n-2}-{\sqrt{(n+1)(n+2)}
\over n+1+\beta-\varepsilon}\,b_{n+2}
\right]  
\end{eqnarray}
and
\begin{eqnarray}
\label{eq14}
\left[n+\beta-\varepsilon-{\tilde{\lambda}_R}^2{n+1\over
n+1+\alpha-\varepsilon}-{\tilde{\lambda}_D}^2{n\over
n-1+\alpha-\varepsilon}\right]a_{n}=
\nonumber\\
\nonumber\\
-i\tilde{\lambda}_R\tilde{\lambda}_D\left[{\sqrt{n(n-1)}
\over n-1+\alpha-\varepsilon}\,a_{n-2}-{\sqrt{(n+1)(n+2)}
\over n+1+\alpha-\varepsilon}\,a_{n+2}
\right] .
\end{eqnarray}
The approximate solution is obtained 
by taking $a_{n-2}=a_{n+2}=b_{n-2}=b_{n+2}=0$ in the above equations. 
This means that for each level $|n\rangle$, the SO interaction is 
allowed to couple it only with the $|n-1\rangle$ and $|n+1\rangle$ levels.
This solution, which
consists of a $|n_d\rangle$ and a $|n_u\rangle$ spinor,
is therefore obtained by solving first
the secular, cubic equation
\begin{equation}
\label{eq15}
(n+\alpha-\varepsilon)(n-1+\beta-\varepsilon)(n+1+\beta-\varepsilon)=
{\tilde{\lambda}_R}^2 n(n+1+\beta-\varepsilon)
+{\tilde{\lambda}_D}^2(n+1)(n-1+\beta-\varepsilon)\;\;\;\; .
\end{equation}  
Together with the equations
\begin{eqnarray}
(n-1+\beta-\varepsilon)a_{n-1}=
-i\tilde{\lambda}_R\sqrt{n}\,b_{n}
\nonumber\\
(n+1+\beta-\varepsilon)a_{n+1}=
-\tilde{\lambda}_D\sqrt{n+1}\,b_{n}
\label{eq16}
\end{eqnarray}  
and the normalization condition $\vert a_{n-1}\vert^2+\vert a_{n+1}\vert^2
+\vert b_{n}\vert^2=1$, they determine the $|n_d\rangle$ solution.
The  solution corresponding to the $|n_u\rangle$ spinor
is obtained by solving the secular equation
\begin{equation}
\label{eq17}
(n+\beta-\varepsilon)(n-1+\alpha-\varepsilon)(n+1+\alpha-\varepsilon)=
{\tilde{\lambda}_R}^2 (n+1)(n-1+\alpha-\varepsilon)
+{\tilde{\lambda}_D}^2n(n+1+\alpha-\varepsilon)\;\;\;\; .
\end{equation}
Together with the equations
\begin{eqnarray}
(n-1+\alpha-\varepsilon)b_{n-1}=
-\tilde{\lambda}_D\sqrt{n}\,a_{n}
\nonumber\\   
(n+1+\alpha-\varepsilon)b_{n+1}=
i\tilde{\lambda}_R\sqrt{n+1}\,a_{n}
\label{eq18}
\end{eqnarray}
and $\vert a_{n}\vert^2+\vert b_{n-1}\vert^2
+\vert b_{n+1}\vert^2=1$, they determine the $|n_d\rangle$ solution.

Since all the estimates available in the literature (see for example Refs.
\onlinecite{Ton04,Kon05,Mal06} and Refs. therein) yield
$\lambda_{R,D}^2$ values of the order of 10 $\mu$eV, and
$\omega_c$ in GaAs is of the order of the meV even at
small $B (\sim 1$ T),  it is worth to  examine the above solutions in the  
${\tilde{\lambda}_{R,D}}^2=2\lambda_{R,D}^2/\omega_c\ll1$ limit, in
which the secular equations 
have solutions easy to interpret.

To order ${\tilde{\lambda}_{R,D}}^2$, the relevant solution to Eq.
(\ref{eq15}) containing both  SO terms is
\begin{equation}
\varepsilon_n^d=n+\alpha + 2n{\lambda^2_{R}\over \omega_c+\omega_L}
- 2(n+1){\lambda^2_{D}\over \omega_c-\omega_L} \;\;\;\; ,
\label{eq19}
\end{equation}
that corresponds to the spinor  $| n_d \rangle$
\begin{equation}
|n_d \rangle =
 \left(\begin{array}{c}
a^{\varepsilon_n^d}_{n-1}\, |n-1 \rangle + a^{\varepsilon_n^d}_{n+1}\, |n+1 \rangle
 \\ 
b^{\varepsilon_n^d}_{n} |n\rangle
 \end{array} 
 \right)
\label{eq19b}
\end{equation}
with coefficients
\begin{eqnarray}
a^{\varepsilon_n^d}_{n-1}&=&
i\tilde{\lambda}_R\sqrt{n}{\omega_c\over \omega_c+\omega_L}
\nonumber\\
a^{\varepsilon_n^d}_{n+1}&=&
-\tilde{\lambda}_D\sqrt{n+1}{\omega_c\over \omega_c-\omega_L}
\nonumber\\
b^{\varepsilon_n^d}_{n}&=&1 - {1\over2}{\tilde{\lambda}_R}^2
n\left({\omega_c\over \omega_c+\omega_L}\right)^2
-{1\over2}{\tilde{\lambda}_D}^2(n+1)\left({\omega_c\over \omega_c-\omega_L}\right)^2
\;\;\; .
\label{eq20}
\end{eqnarray}
In the following, we will
refer to this solution as to the 
quasi spin-down (qdown) solution, since
in the zero spin-orbit coupling limit 
$|n_d \rangle$ becomes
$|n\rangle$$\left(\begin{array}{c} \stackrel{\textstyle 0}{1}
\end{array} \right)$.
Analogously, Eq. (\ref{eq17}) has the solution
\begin{equation}
\varepsilon_n^u=n+\beta - 2(n+1){\lambda^2_{R}\over 
\omega_c+\omega_L}
+ 2 n {\lambda^2_{D}\over \omega_c-\omega_L} ~~,
\label{eq21}
\end{equation}
that corresponds to the spinor  $|n_u \rangle$
\begin{equation}
|n_u \rangle =
 \left(\begin{array}{c}
a^{\varepsilon_n^u}_n \, |n \rangle  \\ 
b^{\varepsilon_n^u}_{n-1}\, |n-1 \rangle + b^{\varepsilon_n^u}_{n+1}\, |n+1 \rangle
 \end{array} 
 \right)
\label{eq21b}
\end{equation}
with coefficients 
\begin{eqnarray}
b^{\varepsilon_n^u}_{n-1}&=&
\tilde{\lambda}_D\sqrt{n}{\omega_c\over \omega_c-\omega_L}
\nonumber\\
b^{\varepsilon_n^u}_{n+1}&=&
i\tilde{\lambda}_R\sqrt{n+1}{\omega_c\over \omega_c+\omega_L}
\nonumber\\
a^{\varepsilon_n^u}_{n}&=&1 - {1\over2}{\tilde{\lambda}_R}^2
(n+1)\left({\omega_c\over \omega_c+\omega_L}\right)^2
-{1\over2}{\tilde{\lambda}_D}^2 n \left({\omega_c\over \omega_c-\omega_L}\right)^2
\;\;\; .
\label{eq22}
\end{eqnarray}
In the following, we will
refer to this solution as to the 
quasi spin-up (qup) solution, since
in the zero spin-orbit coupling limit 
$|n_u \rangle$ becomes
$|n\rangle$$\left(\begin{array}{c} \stackrel{\textstyle 1}{0}
\end{array} \right)$. When either $\lambda_R$ or $\lambda_D$ are 
zero, Eqs. (\ref{eq19b}) and (\ref{eq21b}) reduce to the exact Eqs.
(\ref{eq8c}) and (\ref{eq10c}), respectively, and the corresponding
$a_i$ and $b_i$
coefficients, valid up to order $\lambda_{R,D}$, can be extracted from
Eqs. (\ref{eq20}) and (\ref{eq22}).
These Eqs. show that $a_n$ and $b_n$ are of order $O(1)$, whereas
$a_{n\pm1}$ and $b_{n\pm1}$ are of order $O(\lambda_{R,D})$, and
$a_{n\pm2}$ and $b_{n\pm2}$ are of order $O(\lambda_{R,D}^2)$.
This shows that the neglected terms in Eqs. (\ref{eq12}) and
(\ref{eq14}) are of order $O(\lambda_{R,D}^4)$.

The sp energies obtained from Eqs. (\ref{eq19}) and
(\ref{eq21}), valid in the $\lambda^2_{R,D}/\omega_c\ll1$ limit,
are
\begin{eqnarray}
E_n^d &=& (n+{1\over2})\omega_c+{\omega_L\over2} + 2n\lambda^2_{R}{\omega_c\over 
\omega_c+\omega_L}
- 2(n+1)\lambda^2_{D}{\omega_c\over \omega_c-\omega_L} ~~
\nonumber\\
E_n^u &=& (n+{1\over2})\omega_c-{\omega_L\over2}  - 2(n+1)\lambda^2_{R}{\omega_c\over 
\omega_c+\omega_L}
+ 2 n \lambda^2_{D}{\omega_c\over \omega_c-\omega_L} \;\;\;\ .
\label{eq23}
\end{eqnarray}
Together with the structure of the associated spinors,
Eqs. (\ref{eq19b}) and (\ref{eq21b}), this sp energy spectrum
constitutes one of the main results of our work.
By suitable differences of these energies, one may obtain the 
sp transition energies discussed in the next Sec.

The above sp energies
coincide with  the ones that can be derived from second order
perturbation theory with the standard expression
\begin{equation}
\label{eq24}
E_n^{(2)}={1\over4}\sum_{m\ne n}{\vert\langle m\vert
-i\tilde{\lambda}_R\,\omega_c(a^+\sigma_- - a^-\sigma_+)
+ \tilde{\lambda}_D\,\omega_c(a^+\sigma_+ + a^-\sigma_-)\vert
n\rangle\vert^2\over E_n^{0}-E_m^{0}} \;\;\;\; ,
\end{equation}
where $|n\rangle= |n,\uparrow\rangle$,
$|n,\downarrow\rangle$
are the spin-up and spin-down eigenstates of the sp Hamiltonian
$\frac{1}{2}(a^+a^- + a^-a^+)\omega_c - {1\over2}\omega_L\sigma_z$
with eigenvalues
$E_n^{0}(\uparrow)=(n+{1\over2})\omega_c-{1\over2}\omega_L$,
and  $E_n^{0}(\downarrow)=(n+{1\over2})\omega_c+{1\over2}\omega_L$, respectively.

The approximate solution Eq. (\ref{eq23}) is very accurate
in the high $B$ limit (see below). It also carries an interesting
information in the opposite limit of vanishing $B$. In this limit
($\omega_L,\omega_c\ll\lambda^2_{R,D})$, 
Eqs. (\ref{eq15}) and (\ref{eq17}) yield the solutions  
\begin{eqnarray}
\label{eq25}
E_n^d&=&\sqrt{2\omega_c[n \lambda^2_{R}+(n+1)\lambda^2_{D}]}
\nonumber\\
\nonumber\\
E_n^u&=&\sqrt{2\omega_c[(n+1)\lambda^2_{R}+n \lambda^2_{D}]}
\end{eqnarray}
which show that, at $B\simeq0$, to order $\lambda^2_{R,D}$ 
the Landau levels are not 
split due to the SO interaction, as one might
naively infer from Eqs. (\ref{eq23}). Another merit of the
approximate solution is that it
displays in a transparent way the interplay
between the three spin-dependent interactions, namely Zeeman,
Rashba and Dresselhaus. 
Such interplay has been also discussed in Ref. \onlinecite{Mal06},
in relation with the violation of the Larmor theorem due to the SO
couplings,
and in Ref. \onlinecite{Val06}, where the Zeeman and SO interplay is discussed
using  the unitarily transformed Hamiltonian technique.
Note also that in GaAs quantum wells, which are the object of
application in this paper, due to the sign of $g^*$,
the lowest energy level is the qup one at the energy
$E_0^u={1\over2}\omega_c-{1\over2}\omega_L - 
2\lambda^2_{R}\,\omega_c/(\omega_c+\omega_L)$,
containing the Rashba contribution alone, whereas the following level
is the qdown one at the energy 
$E_0^d={1\over2}\omega_c+{1\over2}\omega_L -
2\lambda^2_{D}\,\omega_c/(\omega_c-\omega_L)$,
containing the Dresselhaus contribution alone. For all the other levels 
both SO terms contribute to the level energies.

\begin{figure}[t] 
\centerline{\includegraphics[width=14cm,clip]{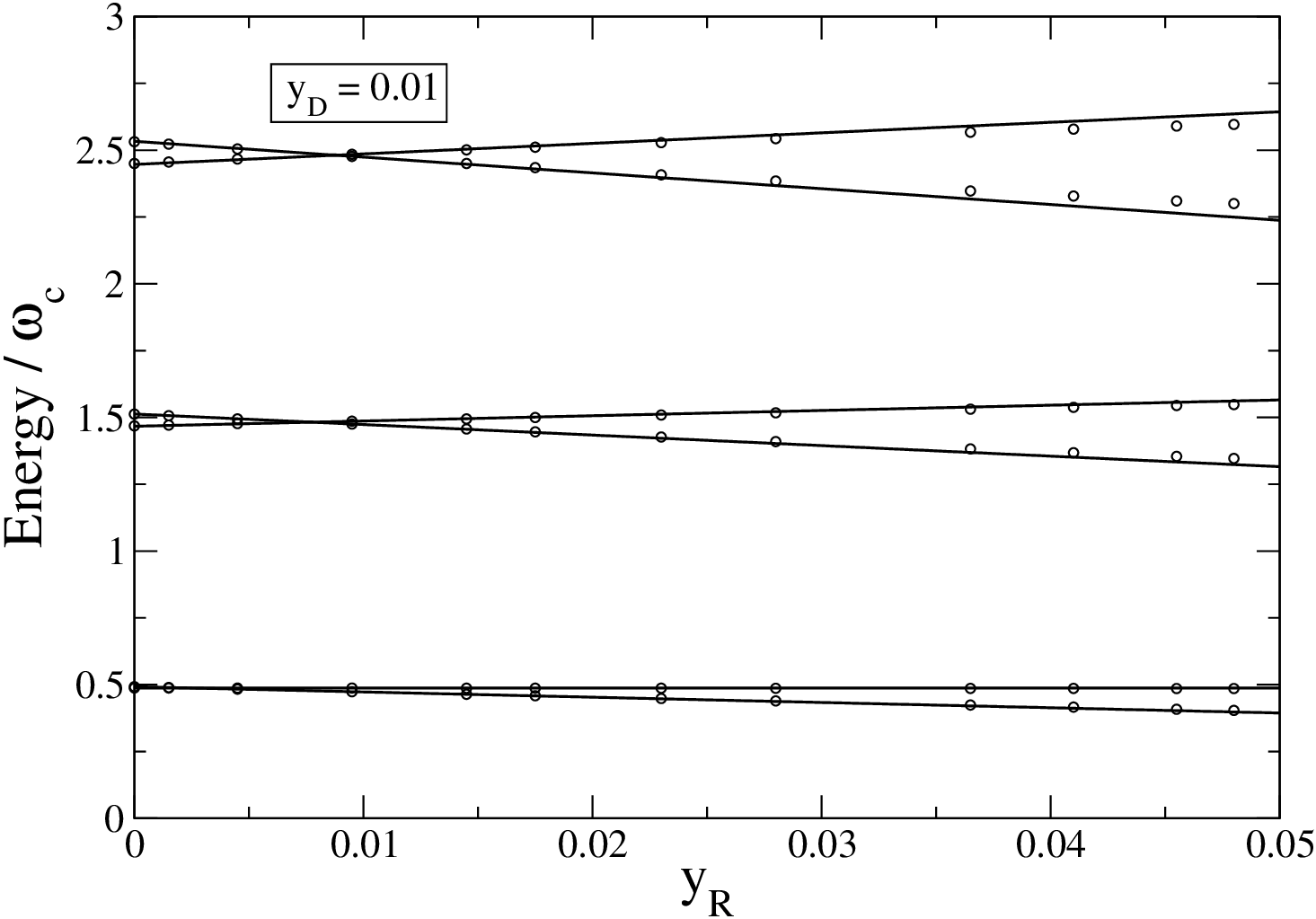}}
\caption{
Lower energy levels for a GaAs quantum well as a function of the Rashba
intensity $y_R =\lambda_{R}^2/\omega_c$ 
 for a fixed Dresselhaus intensity
$y_D = \lambda_{D}^2/\omega_c=0.01$.
 Solid
lines are the analytical result, Eq.\ (\ref{eq23}), while
symbols correspond to the exact diagonalization, Eq.\ (\ref{exd}).
} 
\label{fig1} 
\end{figure}

We have assessed the quality of the above analytical solutions, Eqs.\ (\ref{eq23}),
by comparing them with exact numerical results for some particular cases.
Indeed, 
the exact solution to Eqs.\ (\ref{eq6}) can be obtained in the truncated space
spanned by the lower ${\cal N}$ oscillator levels. Mathematically, Eqs.\ (\ref{eq6})
are then cast into a linear eigenvalue problem of the type
\begin{equation}
\label{exd}
{\bf M} 
\left(\begin{array}{c}
\stackrel{\textstyle {\bf a}}{\textstyle {\bf b}} \end{array} \right)
= \varepsilon
\left(\begin{array}{c}
\stackrel{\textstyle {\bf a}}{\textstyle {\bf b}} \end{array} \right)
\;\;\;\; ,
\end{equation}
where ${\bf M}$ is a $2{\cal N}\times2{\cal N}$ matrix, while ${\bf a}$ and
${\bf b}$ are column vectors made with the sets of coefficients 
$\{a_n, n=0,\dots, {\cal N}-1\}$ and $\{b_n, n=0,\dots, {\cal N}-1\}$, respectively.
We have diagonalized  ${\bf M}$ using a large enough ${\cal N}$
to ensure good convergence in the lower eigenvalues. Fig.\ \ref{fig1} displays
a comparison of numerical (symbols) and analytical (solid lines)
energies as a function of the Rashba SO strength, for a fixed 
Dresselhaus strength, both in units of $\omega_c$, namely
$y_R =\lambda_{R}^2/\omega_c$ and  $y_D = \lambda_{D}^2/\omega_c= 0.01$.
The chosen values for $y_D$ and $y_R$ are within the
expected range for a GaAs quantum well. For instance, if
$m \lambda^2_{R,D}/\hbar^2 \sim 10 \mu$eV and $B \sim 1$T,
$(m \lambda^2_{R,D}/\hbar^2)/(\hbar \omega_c)\sim 10^{-2}$.
There is an excellent agreement between 
analytical and numerical results, differences starting to be visible
only for strong Rashba intensities
and high Landau bands. Actually, 
in Fig.\ \ref{fig1} the largest value of the adimensional ratio
between Rashba SO and cyclotron energy  
$y_R =\lambda_{R}^2/\omega_c$ is 0.05, 
small enough to validate the analytical expression. Notice, 
however, that for larger $y_R$ values -not shown in the figure- i.e., 
for small enough $B$, Eqs.\ (\ref{eq23}) no longer reproduces the
numerical results. For GaAs this happens
for magnetic fields below 0.1 T.
Similarly, Fig. \ref{fig2}  displays
a comparison of numerical (symbols) and analytical (solid lines)
energies as a function of the Dresselhaus SO strength $y_D$, for a fixed
Rashba strength $y_R = \lambda_{D}^2/\omega_c= 0.01$.
For every Landau level, both figures show a crossing between
the $|n_u\rangle$ state, which is at lower energy for 
$y_{R,D} \ll 0.01$ because $g^* <0$,  and the $|n_d\rangle$ state,
that eventually lies lower in energy. This crossing is due to the
interplay between both SO terms.

\section{Single-particle level transitions induced by applied 
electromagnetic fields.}

We can use the preceding results to study the sp transitions induced
in the system by the interaction with a
left-circular polarized electromagnetic wave 
propagating along the $z$-direction, i.e., perpendicular to
the plane of motion of the electrons,
whose vector potential is ${\bf A}(t)=2A(\cos\theta \hat i+ \sin\theta
\hat j)$, with $\theta=\omega t - q z$. The sp
interaction Hamiltonian  ${\bf J}\cdot{\bf A}/c +
g^* \mu_B\, {\bf s} \cdot (\nabla \times {\bf A})$, where
${\bf J}=e\,{\bf v}/\sqrt{\epsilon}$, reads
\begin{equation}
\label{eq26}
h_{int}={e\over c\sqrt{\epsilon}}A\left(v_- e^{i\theta}+v_+ e^{-i\theta}\right)
+{1\over2}g^*\mu_B q A\left(\sigma_- e^{i\theta}+\sigma_+ e^{-i\theta}\right)
\;\;\;\; , 
\end{equation}
where the velocity operator $v_\pm$ is
defined as $v_\pm \equiv -i[x\pm i y, H] =
P^{\pm}\pm i\lambda_R \sigma_\pm + \lambda_D\sigma_\mp$.

The Hamiltonian $h_{int}$ can be rewritten as
\begin{equation}
\label{eq271}
h_{int}={e\over c\sqrt{\epsilon}}A\sqrt{2\omega_c}\left(\alpha^-
e^{i\theta}+\alpha^+ e^{-i\theta}\right)
+{1\over2}g^*\mu_B q A\left(\sigma_- e^{i\theta}+\sigma_+
e^{-i\theta}\right) \;\;\;\; ,
\end{equation}
where the operators $\alpha^+$ and $\alpha^-$ acting on the spinor
$|\phi \rangle$ are
\begin{equation}
\alpha^+=\left[ \begin{array}{cc}
a^+ & i\tilde{\lambda}_R\\\tilde{\lambda}_D & a^+\end{array} \right]~~,
~~\alpha^-=\left[ \begin{array}{cc}
a^- & \tilde{\lambda}_D\\-i\tilde{\lambda}_R & a^-\end{array} \right]~~.
\label{eq28}
\end{equation}
In the dipole approximation ($q \approx 0$),
the charge-density excitation operator is  $v_\pm$. We note
that, even in the presence of e-e interactions,
this operator satisfies the f-sum rule:
\begin{equation}
\label{eq29}
\sum_n\langle0\vert x\mp i y\vert n \rangle\langle n\vert -iv_\pm\vert0\rangle=
\sum_n\omega_{n0}\vert\langle n\vert x\pm i y\vert0\rangle\vert^2=
{1\over2}\langle0\vert[x\mp i y,[H,x\pm i y]]\vert0\rangle=2 N~~,
\end{equation}
where  $N$ is the electron number and $\omega_{n0}$ are the excitation
energies. 

We consider next several useful examples of sp matrix elements
involving the operators $\alpha^+$, which is proportional to
$v_+$, and $\sigma_-$, and the qup and qdown sp states of Eqs.
(\ref{eq19b}) and (\ref{eq21b}).
For the operator $\alpha^+$, we can write in general
\begin{equation}
\langle \psi |\alpha^+| \phi \rangle  =
\psi^*_1a^+\phi_1 +i\tilde{\lambda}_R\psi^*_1\phi_2 
+\tilde{\lambda}_D\psi^*_2\phi_1 +\psi^*_2a^+\phi_2
\;\;\;\; ,
\label{eq30}
\end{equation}
and have to distinguish between
qup-qup,  qdown-qdown, qup-qdown, and qdown-qup transitions.
The qup-qup
and qdown-qdown transitions represent
the usual CR, and the
qup-qdown and qdown-qup are related to spin-flip transitions.

Let us start with the qup-qup and qdown-qdown  transitions.
To the order $\lambda^2_{R,D}$ they
are dominated by the  transition $n\rightarrow n+1$ at the energies
$E_{n+1}^d-E_{n}^d$
and 
$E_{n+1}^u-E_{n}^u$
with  matrix elements 
$|\langle (n+1)_u\vert \alpha^+\vert n_u\rangle| =
|\langle (n+1)_d\vert \alpha^+\vert n_d\rangle| =\sqrt{n+1}$.
The energy splitting of the cyclotron resonance
is 
\begin{equation}
\label{eq33}
\Delta E_{CR} = \left\vert4\lambda^2_{R}{\omega_c\over
\omega_c+\omega_L}
-4\lambda^2_{D}{\omega_c\over \omega_c-\omega_L}\right\vert ~~.
\end{equation}
The $\alpha^+$ excitation operator also induces
a qup-qdown transition with  energy
$E_{n}^d-E_{n}^u$
and matrix element 
$|\langle n_{d}\vert \alpha^+\vert n_{u}\rangle|
=\tilde{\lambda}_D\,\omega_L/(\omega_c-\omega_L)$.
This is a spin-flip transition.
In particular, when $n=0$ it is related to the Larmor
resonance at the energy\cite{Mal06}
\begin{equation}
\label{eq333}
\Delta E_{L} = \omega_L + 2\left(\lambda^2_{R}{\omega_c\over
\omega_c+\omega_L}
-\lambda^2_{D}{\omega_c\over \omega_c-\omega_L}\right) ~~.
\end{equation}
Note that the transition matrix 
element 
is linear in $\tilde{\lambda}_D$, and
that in the presence of the Rashba interaction alone,
$\alpha^+$ causes no spin-flip transition.

For the operator $\sigma_-$ one gets 
$\langle \psi |\sigma_- |\phi \rangle = 2\psi^*_2\phi_1$.
The dominant transition is the spin-flip excitation at energy
$E_{n}^d-E_{n}^u$ with
matrix element $|\langle n_{d}\vert \sigma_-\vert n_{u}\rangle|=2$.
The qup-qup and
qdown-qdown cyclotron resonances at energies
$E_{n+1}^d-E_{n}^d$ and $E_{n+1}^u-E_{n}^u$
are also excited with 
strengths  $|\langle (n+1)_u\vert \sigma_-\vert n_u\rangle|=
|\langle (n+1)_d\vert \sigma_-\vert n_d\rangle|=
2\tilde{\lambda}_D\sqrt{n+1}\,\omega_c/(\omega_c-\omega_L)$.

Other excitations that deserve some  attention are those
induced by the operators
$\alpha^+\sigma_\pm$ and  $\alpha^+\sigma_z$. 
They are detected in inelastic
light scattering experiments as spin dipole resonances.\cite{Eri99}
The operator $\alpha^+\sigma_z$ excites the same cyclotron
states as $\alpha^+$,  at the energies
$E_{n+1}^d-E_{n}^d$ and $E_{n+1}^u-E_{n}^u$, and  with
the same transition matrix element $\sqrt{n+1}$. 
In contrast, the operator
$\alpha^+\sigma_+$ mainly induces the transition from  qdown
to qup states at the energy  $E_{n+1}^u-E_{n}^d$,
whereas the operator
$\alpha^+\sigma_-$ induces the transition from  qup
to qdown states at the energy $E_{n+1}^d-E_{n}^u$.
The transition matrix elements  are given by
$|\langle (n+1)_u\vert \alpha^+\sigma_+\vert n_d\rangle| =
|\langle (n+1)_d\vert \alpha^+\sigma_-\vert n_u\rangle|
=2\sqrt{n+1}$.
We  thus see that the dipole transitions between Landau levels
$|n\rangle$ and $|n+1\rangle$ at `unperturbed' energies
$E_{n+1}-E_{n}$ are split  by the SO interaction,
an effect that under some circumstances may be observed,
as will be discussed in Sec. V.

\section{Electron-electron interaction and sum rules}

In this section our aim is to discuss the role played by 
the e-e interaction in  the physical
processes in which SO effects can be  important
and, as a consequence, have a chance 
to be  experimentally detected. 
Since we have obtained a   spinor
basis that includes the SO effects (SO basis), one might
use it to diagonalize the 
Coulomb interaction. This has been done, for
example, in Ref. \onlinecite{Cal05}, 
where the spinor basis
Eq. (\ref{eq8c})  has been used to
study the influence of the Rashba interaction 
on the incompressible Laughlin state. 
One could also
use the SO basis Eqs. (\ref{eq19b}) and (\ref{eq21b})
to solve the random-phase-approximation
equations,\cite{Kal84} or 
to study SO effects on the collective states
of the quantum well  
in the adiabatic time-dependent
local-spin-current density approximation.\cite{Mal06,Ser99}
We have chosen a different way to incorporate interaction
effects that, while being more approximate,
it is accurate enough and allows one to obtain simple
analytical expressions for the quantities of interest here.
It is  the sum rule approach, which is well suited
to address the interplay between SO coupling and the
e-e interaction in some relevant excitation processes. 
 
Let us firstly recall
that in the absence of the SO coupling,
two important theorems hold for the  quantum well Hamiltonian $H$, 
in which the e-e interaction is included.
They are the Kohn theorem
\begin{equation}
\label{eq38}
[H,\sum_j P^+_j]=\omega_c\sum_j P^+_j ~~,
\end{equation}
which tells  us that, in photoabsorption
experiments on quantum wells,
a narrow absorption peak must appear at the cyclotron
frequency $\omega=\omega_c$ excited by the cyclotron operator
$\sum_j P^+_j$, and the Larmor theorem
\begin{equation}
\label{eq39}
[H,S_-]=\omega_L S_-~~,
\end{equation}
which states that in inelastic light scattering experiments at small
transferred momentum, or in electron-spin resonance experiments,
a narrow collective state 
must be excited by the Larmor operator $S_-=\sum_j \sigma^j_-$
at the Larmor frequency.
These two modes are not  influenced by the e-e interaction.
Things radically change if we include in $H$ the  SO interaction. We then obtain
\begin{equation}
\label{eq40}
[H,\sum_j P^+_j]=\omega_c\sum_j\left( P^+ + i\lambda_R\sigma_+  
+\lambda_D\sigma_-\right)_j
\end{equation}
and
\begin{equation}
\label{eq41}
[H,S_-]=\omega_L S_- +\sum_j\left(2i\lambda_R P^-\sigma_z  
+2\lambda_D P^+\sigma_z\right)_j \;\;\;\; ,
\end{equation}
which show that the SO interaction couples the cyclotron (dipole and
spin
dipole) and Larmor modes. We have studied in Sec III the effects of the
SO coupling on these excitations in the absence of the Coulomb interaction. 
Now, we want to determine whether
the presence of both the SO and Coulomb interactions 
has an effect on the Larmor and cyclotron frequencies or, on the contrary, the
results of the previous Sec. still hold. 
With this goal in mind, we introduce the
following mixed sum rules\cite{Physrep,Lip04}
\begin{eqnarray}
m_k^\pm&=&{1\over2} 
\sum_n\omega^k_{n0}\left(\langle0\vert F\vert \phi_n\rangle\langle
\phi_n\vert G^{\dagger}\vert 0\rangle
\pm \langle0\vert G^{\dagger}\vert \phi_n\rangle\langle \phi_n\vert
F\vert 0\rangle\right) \nonumber\\
&=&{1\over2}\left(\langle0\vert F(H-E_0)^k G^{\dagger}\vert
0\rangle\pm\langle0 \vert G^{\dagger}(H-E_0)^k F\vert 0\rangle\right)
\; ,
\label{eq42}
\end{eqnarray}
where $|0\rangle$ and $|\phi_n\rangle$ are the exact  gs and 
excited states of the full Hamiltonian $H$ (including e-e
interactions), and $\omega_{n0}=E_n-E_0$ are the corresponding
excitation energies. For $k=0-3$ we obtain
\begin{eqnarray}
m_0^-&=&{1\over2}\langle0\vert [F, G^{\dagger}]\vert 0\rangle
\nonumber\\
m_1^+&=&{1\over2}\langle0\vert [F,[H, G^{\dagger}]]\vert 0\rangle
\nonumber\\
m_2^-&=&{1\over2}\langle0\vert [[F,H],[H, G^{\dagger}]]\vert 0\rangle
\nonumber\\
m_3^+&=&{1\over2}\langle0\vert [[F,H],[H,[H, G^{\dagger}]]]\vert
0\rangle 
\;\;\;\;\;  .
\label{eq43}
\end{eqnarray} 
Clearly, the
more sum rules are known, the better knowledge of the Hamiltonian spectrum.
With the four sum rules of Eq. (\ref{eq43})
we can obtain information only on two excited states -see below.
Consequently,
we will limit the analysis to the cases in which either the Rashba or
Dresselhaus SO terms are present because, as one can  see from
Eq. (\ref{eq40}) and Eq. (\ref{eq41}) as well, in this case  only
two states would then be coupled by the corresponding  SO
interaction.

Let us first consider the case where  $F=G=\sum_i P^-_i$, i.e.,
$G^{\dagger}$ is the cyclotron operator. Evaluating
the commutators in Eqs. (\ref{eq43}) we have, to order
$\lambda_{R,D}^2$,
\begin{eqnarray}
m^-_0 &=& 2N\omega_c
\nonumber\\
m^+_1 &=& 2N\omega^2_c
\nonumber\\
m^-_2 &=&  2N\omega^3_c \left[1-{2\over\omega_c}(\lambda^2_{R} -
\lambda^2_{D})\right]
\nonumber\\
m^+_3 &=& 2N\omega^4_c \left[1-{4\over\omega_c}(\lambda^2_{R} -
\lambda^2_{D})+{2\omega_L\over\omega^2_c}(\lambda^2_{R} +
\lambda^2_{D})\right] \;\;\;\; .
\label{eq44}
\end{eqnarray}
To obtain these Eqs.  we have used 
that $\sum_i P^-_i\vert 0\rangle=0$ and  have assumed that the gs
of the system is fully polarized, i.e.,
$\langle 0\vert \sum_i \sigma^i_z\vert 0\rangle=N$.
As such,
these expressions are useless unless the left-hand side can be
directly evaluated from the definition Eq. (\ref{eq42}), and
this evaluation
yields a closed expression for the excitation energies and transition
matrix elements. This is the case if we consider   either of
the $\lambda_{R,D}$ terms alone, because only two states
are excited by the cyclotron operator $G^{\dagger}= \sum_i P^+_i$
acting on the gs $|0\rangle$.
Dropping e.g. the $\lambda_R$ term, a straightforward calculation yields
\begin{eqnarray}
\pi_1 &=& 2N\omega_c\left[1 -
{2\omega_c\over(\omega_c-\omega_L)^2}\lambda^2_{D}\right] \nonumber\\
\pi_2 &=& 2N {2\omega_c^2\over(\omega_c-\omega_L)^2}\lambda^2_{D}
\nonumber\\
\omega_{10} &=& \omega_c + {2\omega_c\over\omega_c-\omega_L}\lambda^2_{D} 
\nonumber\\
\omega_{20} &=& \omega_L - {2\omega_c\over\omega_c-\omega_L}\lambda^2_{D}
\;\;\;\; ,
\label{eq45}
\end{eqnarray}  
where $\pi_1$ and  $\pi_2$ are the transition strengths
to the cyclotron 
$\vert \phi_{n_1}\rangle$ and Larmor $\vert \phi_{n_2}\rangle$ states,
$\pi_1=\vert \langle \phi_{n_1}\vert \sum_i P^+_i\vert 0\rangle\vert^2$
and
$\pi_2=\vert \langle \phi_{n_2}\vert \sum_i P^+_i\vert 0\rangle\vert^2$,
and $\omega_{10}$, $\omega_{20}$ are the respective excitation
energies.
This is in full agreement with the results of  Sec. III, and
shows that the e-e interaction does not
affect the frequency and transition strengths of the cyclotron
and Larmor resonances. 

The case $\lambda_{D}=0$ can be worked out similarly, and the same
conclusion may be extracted. We recall and stress again the results
obtained in the previous section, namely that when $\lambda_{D}=0$,
the Larmor state $\vert \phi_{n_2}\rangle$ is not excited
by the cyclotron operator $\sum_i P^+_i$ ( $\pi_2$ turns out to be
zero).
Alternatively, all previous calculations could have been carried out
using for $G^{\dagger}$ the Larmor operator, namely,
$F=G=\sum_i\sigma^i_+$.
Assuming again that $\langle 0\vert \sum_i \sigma^i_z\vert 0\rangle=N$,
we obtain the same results and draw the same conclusions as before. This
is a consequence of the structure of Eqs. (\ref{eq40}) and (\ref{eq41}).

Using more sum rules, e.g. $m^-_4$ and $m^+_5$, one may
obtain information on other states that can be excited by the 
cyclotron operator $\sum_i P^+_i$. Their consideration 
shows that the e-e interaction does not
affect, to order $\lambda^2_{R,D}$,  neither the cyclotron nor the
Larmor state, whose frequencies are the same as determined in
Sec. III when both the SO terms are included in $H$.

When the gs of the system has
both qup and qdown occupied states,\cite{Ton04}
the spin dipole operator $\sum_i P^+_i\sigma^i_z$
entering Eq. (\ref{eq41}) excites a state at an energy
$\omega_c(1+{\cal K})$  -see below, instead of
$\omega_c$ as it
corresponds to the cyclotron (charge dipole) operator
$\sum_i P^+_i$, and the results in Eq. (\ref{eq45}) 
must be corrected for. This effect is not related to
the SO interaction, and appears even in the absence of it.
The spin dipole operator does not commute with
the e-e interaction as the
cyclotron operator $\sum_i P^+_i$ does, and ${\cal K}$
is precisely the contribution to the spin dipole operator
$m^+_1$ sum rule
arising from the e-e interaction when one takes
$F=G=\sum_i P^-_i\sigma^i_z$:
\begin{eqnarray}
m_1^{+} &=&
\sum_n\omega_{n0}\vert\langle \phi_n\vert \sum_i P^+_i\sigma^i_z\vert
0\rangle\vert^2 \nonumber\\
&=&{1\over2}\langle0\vert [\sum_j P^-_j\sigma^j_z,[H, \sum_i
P^+_i\sigma^i_z]]\vert 0\rangle=N \omega_c^2(1+{\cal K}) \;\;\;\; ,
\label{eq46}
\end{eqnarray}
where
\begin{equation}
\label{eq47}
{\cal K}={1\over 2N \omega_c^2}\langle0\vert \sum_{i<j}{\bf\nabla}^2_{r_{ij}}
V(r_{ij})(\sigma^i_z-\sigma^j_z)^2\vert0\rangle~~.
\end{equation}
${\cal K}$ can be extracted from  inelastic light 
scattering experiments.\cite{Eri99} It turns out to be zero
for fully polarized ground states, and small and negative
-of the order of 10$^{-2}$- otherwise. Similarly, the
spin flip dipole operators $\sum_i P^+_i\sigma^i_\pm$,
whose excitations can be also measured by inelastic light scattering,
do not commute with the e-e interaction, which give rise to
some energy corrections. It turns out that
these corrections are equal for the three spin
dipole operators $\sum_i P^+_i\sigma^i_{z,\pm}$ 
because the value of ${\cal K}$ is the same for all them.
Hence, the energy splittings
among these excitations are not influenced by the e-e interaction,
depending only on the Zeeman  and SO energies as found
and discussed at the end of  Sec III.

Finally, we want to comment on the consequences of the
failure of the Kohn  theorem due to the
SO coupling using the $m_1^{+}$ sum rule for 
$F=\sum_i P^-_i\sigma^i_z$  and $G=\sum_i P^-_i$:
\begin{eqnarray}
m_1^{+} &=&
\sum_n\omega_{n0}\langle0\vert \sum_i P^-_i\sigma^i_z\vert n\rangle\langle n\vert 
\sum_j P^+_j\vert 0\rangle
\nonumber\\
&=&{1\over2}\langle0\vert [\sum_i P^-_i\sigma^i_z,[H, \sum_j
P^+_j]]\vert 0\rangle=\omega_c\langle 0\vert \sum_i \sigma^i_z\vert 0\rangle
\;\;\;\; .
\label{eq48}
\end{eqnarray}  
This sum rule allows to study the interplay between 
charge and spin modes. If we cast it into a sum
over `spin dipole states' $\vert m\rangle$ and another
over `charge dipole states' $\vert \rho\rangle$, we obtain
\begin{eqnarray}
m_1^{+} &=&
\sum_\rho\omega_{\rho0}\langle0\vert \sum_i P^-_i\sigma^i_z\vert \rho\rangle\langle 
\rho\vert\sum_j P^+_j\vert 0\rangle
\nonumber\\
&+&\sum_m\omega_{m0}\langle0\vert \sum_i P^-_i\sigma^i_z
\vert m\rangle\langle m\vert\sum_j P^+_j\vert 0\rangle
\;\;\;\; .
\label{eq49}
\end{eqnarray}
If there is no SO coupling, Kohn's theorem holds, implying that
$\langle m\vert\sum_i P^+_i\vert 0\rangle=0$. Thus, 
when the spin  gs $2S_z=\langle 0\vert \sum_i \sigma^i_z\vert 0\rangle$ is
not zero [otherwise, $m^+_1=0$ from Eq. (\ref{eq48})], only the density
modes
would contribute to $m_1^{+}$ through the $\rho$-sum in Eq. (\ref{eq49}).
On the contrary, if the SO coupling is taken into account, Kohn's theorem 
is violated and the spin and charge dipole states are coupled
to order $\lambda_{R,D}^2$, with $\langle m\vert\sum_i P^+_i\vert 0\rangle$
being now different from zero. 

To be more quantitative, let us assume that only one charge dipole
state, the cyclotron state $\vert \rho\rangle$ at energy
$E_1=\omega_c+O(\lambda^2_{R,D})$,
contributes to the first sum in Eq. (\ref{eq49}), and only one
spin dipole state $\vert m\rangle$, at energy
$E_2=\omega_c(1+{\cal K})+O(\lambda^2_{R,D})$,
contributes  to the second sum, where we have indicated
by $O(\lambda^2_{R,D})$ the SO correction to the cyclotron and spin
dipole energies. Let us define the mixed  strengths
\begin{eqnarray}
\pi_1 &=& \langle0\vert \sum_i P^-_i\sigma^i_z\vert \rho\rangle\langle
\rho\vert\sum_j P^+_j\vert 0\rangle
 \nonumber\\
\pi_2 &=& \langle0\vert \sum_i P^-_i\sigma^i_z\vert m\rangle
\langle m\vert\sum_j P^+_j\vert 0\rangle
\;\;\;\; .
\label{eq491}
\end{eqnarray}
Evaluating the sum rules $m_0^-$ and $m_1^+$  for
the operators $G=\sum_i P^-_i$ and $F=\sum_i P^-_i\sigma^i_z$, one
easily obtains
\begin{eqnarray}
\pi_1 &=& \omega_c\langle 0\vert \sum_i \sigma^i_z\vert 0\rangle{E_2-\omega_c\over E_2-E_1}=
\omega_c\langle 0\vert \sum_i \sigma^i_z\vert 0
\rangle\left(1-{O(\lambda^2_{R,D})\over|\omega_c\,{\cal K}|}\right)\; ,
 \nonumber\\
\pi_2 &=& \omega_c\langle 0\vert \sum_i \sigma^i_z\vert 0\rangle{\omega_c-E_1\over E_2-E_1}=
\omega_c\langle 0\vert \sum_i \sigma^i_z\vert 0\rangle 
{O(\lambda^2_{R,D})\over|\omega_c\,{\cal K}|}
\;\;\;\; .
\label{eq492}
\end{eqnarray}
Eqs. (\ref{eq492}) explicitly show that if $\langle 0\vert \sum_i \sigma^i_z\vert
0\rangle=0$,
or if the SO coupling is neglected,  the mixed
strength $\pi_2$ is zero, and the spin dipole state cannot be excited
in photoabsorption experiments.  
The strength $\pi_2$ is nonzero only at odd filling factors $\nu$
($\nu=2\pi\ell^2 n_e$, where $\ell=\sqrt{\hbar c/e B}$
is the magnetic length and $n_e$ is the electron density),
for which $2 S_z/N=1/\nu$. Besides, when the system is fully polarized
at $\nu=1$, the operators $\sum_i P^-_i$ and $\sum_i P^-_i\sigma^i_z$
coincide and excite the same mode, so there is no splitting.
The SO corrections $O(\lambda^2)$ can be calculated
by taking into account the occupation of the ground state, either using
the sum rule approach of this Section, or the method of unitarily
transforming the Hamiltonian,
as described in Refs. \onlinecite{Ton04,Val06}. This calculation yields
the energy splitting of the CR we discuss in the next Section.

\section{Comparison with experiments and discussion}

An actual confrontation of the theoretical results we
have obtained with the experiments is not an easy task because of
the smallness of the SO effects, and because of the way they are
presented in the available literature, which
makes it extremely difficult to carry out
a quantitative analysis of such a subtle effect. Thus, we
have to satisfy ourselves with a semi-quantitative analysis, or to point 
out that these results are compatible with fairly rough estimated
values of the SO coupling constants. We present now three such examples
and a possible way to increase SO effects so that they could
be easier to determine.

Using unpolarized far-infrared radiation, Manger et
al.\cite{Man01} have measured the cyclotron resonance in 
GaAs quantum wells at different electron densities. The main finding
of the experiment is a well resolved splitting of the CR for  $\nu$=3, 5, and
7, and no significant splitting for $\nu=1$ and for even filling factors.
We have seen that the SO interaction
couples charge-density and spin-density excitations yielding
the SO splitting of the CR given in Eq. (\ref{eq33}).
However, this expression, by itself,  is unable to explain the filling
factor dependence of the observed splitting, for which
one has to bear in mind  that the SO coupling between the
$\sum_i P^-_i$ and $\sum_i P^-_i\sigma^i_z$ operators is strongly
enhanced when the spin gs is not zero,
as explicitly shown in Eq. (\ref{eq492}). 
We have also noted that ${\cal K}$ contributes to the splitting.
Eq. (\ref{eq33}) has to be generalized to include these features.
We obtain
\begin{equation}
\label{eq50}
\Delta E_{CR} = \left\vert{2S_z\over
N}\,4\left(\lambda^2_{R}{\omega_c\over \omega_c+\omega_L}
-\lambda^2_{D}{\omega_c\over \omega_c-\omega_L}\right) + 
{\cal K}\omega_c\right\vert \;\;\;\; ,
\end{equation}
where the factor $2S_z/N$ takes into account the actual sp contents
of the gs. This equation, together with Eq. (\ref{eq492}),
embodies the theoretical explanation of the experimental 
findings.\cite{Man01} In particular,
it gives an appreciable splitting only for odd filling factors, for
which the  spin ground state $S_z$ is not zero.
The analysis of the experimental splittings using the Eq. (\ref{eq50}) 
yields values for the quantity 
$m\vert\lambda^2_{R}-\lambda^2_{D}\vert/\hbar^2$ of about $30\mu$eV,
in agreement with the ones recently used to reproduce the spin splitting
in quantum dots\cite{Kon05} and wells.\cite{Mal06} This is, in our
opinion, one of the most clear evidences of a crucial SO effect
on a physical observable, because its absence would imply that
the physical effect does not show up.

The spin splitting of the first three Landau levels 
of a GaAs quantum  well has been measured 
in a magnetoresistivity experiment by Dobers et al.\cite{Dob88}
We have shown in the previous sections that this splitting
is not influenced by the e-e interaction, and that there is no
spin splitting as $B$ goes to zero 
[Eq. (\ref{eq25})]. Both facts are in agreement
with the analysis of the experimental data, and with
previous theoretical considerations\cite{Mal86} about the $B$-dependence
of the gyromagnetic factor $g^*$, whose determination
was the physical motivation of the magnetoresistivity 
experiment presented in Ref. \onlinecite{Dob88}.
These authors have derived a $B$- and $n$- dependent $g^*$ factor
$g^*(B,n)= g^*_0 -c(n+\frac{1}{2})B$,
where $g^*_0$ and $c$ are fitting constants that depend on the
actual quantum well. The possibility of a SO shift was not considered,
and their chosen law for $g^*$ implies that the spin splitting energy
$\Delta E_n$ does depend on the Landau level index $n$ entering in a
$B^2$ term, as  they have $\Delta E_n=|g^*\mu_B B|$. A $B$ dependence in
$g^*$ is crucial to explain the experimental data, and also to
reproduce them theoretically.\cite{Mal06}

For the spin splitting of the Landau levels we obtain 
\begin{equation}
\label{eq51}
\Delta E_n = \omega_L+2(2n+1)\left(\lambda^2_{R}{\omega_c\over
\omega_c+\omega_L} 
-\lambda^2_{D}{\omega_c\over \omega_c-\omega_L}\right) 
\end{equation}
-recall that $\omega_L=\vert g^*\mu_B B\vert$-
i.e., a splitting that increases with $n$ because of the SO coupling.
This SO correction has been
worked out for the $n=0$ level in Ref. \onlinecite{Mal06} using the
equation of motion method. It is known that the experimental
results\cite{Dob88} for $n=1$ and 2 can be reproduced
if $g^*$ depends on $n$ and $B$, as already shown in that
reference. We have verified that the $n$-dependence of $g^*$ cannot be
mimicked by the $n$-dependence introduced by the SO interaction,
Eq. (\ref{eq51}).
Recently, the analysis of $g^*$ has been extended to
a wider magnetic field range using time-resolved Faraday rotation 
spectroscopy.\cite{Sal01,Sih04}

As a third example, we address the inelastic light scattering 
excitation of the spin dipole  modes at $\nu=2$
as measured in Ref. \onlinecite{Eri99}.
For this filling factor, in the absence of SO coupling
the spin-density inter-Landau level 
spectrum is expected to be a triplet mode\cite{Kal84,Lip99}
excited by the three operators $\sum_i P^+_i\sigma^i_{z,\pm}$
with energy splittings
given by the Zeeman energy $\omega_L$. In the presence of SO interactions,
we still expect a triplet mode to appear. Indeed, for $\nu=2$ we have $S_z=0$ and
the cyclotron and  spin dipole
modes excited by the operator $\sum_i P^+_i\sigma^i_z$ are decoupled, as
previously discussed. Thus,
for this operator only one  single mode should be detected at an average
energy $\omega=\omega_c(1+{\cal K})$.
The other operators $\sum_i P^+_i\sigma^i_{\pm}$,
yield the two other spin dipole modes at the energies
\begin{equation}
\label{eq52}
E^\pm =\omega_c(1+{\cal K}) \pm \omega_L \pm 4\left(\lambda^2_{R}{\omega_c\over
\omega_c+\omega_L}
-\lambda^2_{D}{\omega_c\over \omega_c-\omega_L}\right) \;\;\;\; .
\end{equation}
The splitting is thus symmetric and depends on the
SO strengths. In the experiment, triplet excitations were observed
in all measured samples up to electron densities corresponding to
$r_s=3.3$ (we recall that $r_s=1/\sqrt{\pi n_e}$).
$B$ was accordingly changed to keep the filling factor at $\nu=2$.
Only one triplet mode spectrum at
$B=2.2$ T was shown. From this spectrum, we infer that
there is space for a $\sim 5-10 \%$ SO effect on the splitting, 
assuming that at this fairly small magnetic field,
$g^*$ is that of bulk  GaAs, $g^*=-0.44$. We have estimated that
$m\vert\lambda^2_{R}-\lambda^2_{D}\vert/\hbar^2 \simeq10\mu eV$,
in line with the previous findings. 
Systematic measurements, especially at high $B$
where the splitting is larger, are called for to allow for a
more quantitative analysis.

Another unequivocal signature of spin-orbit effects in quantum wells 
would be the detection of the Larmor state in photoabsorption 
experiments. The strength of this transition is given by
$2{\lambda^2_D\over\omega_c}({\omega_L\over\omega_c-\omega_L})^2$ 
and only depends on the Dresselhaus SO coupling 
-see the comment immediately after Eq. (\ref{eq33}).
In most experiments, $B$ is perpendicularly applied
to the plane of motion of the electrons, and for GaAs
the strength is so small that it has never been resolved.

We finally discuss the effect of tilting the applied magnetic field
using the expressions derived in the Appendix.
Eq. (\ref{eqa3}) can be used to obtain the
splitting of the cyclotron resonance which generalizes
Eq. (\ref{eq33}) for tilted magnetic fields:
\begin{equation}
\Delta E_{CR}=
4\left[ (C_R~{\cal V} +C_D~{\cal Z})
\frac{1}{1 + |g^*| m^* {\cal S}/2} -
(C_R~{\cal Z}+C_D~{\cal V})
\frac{1}{1 - |g^*| m^* {\cal S}/2}
\right] \;\;\;\; ,
\label{eq56}
\end{equation}
where   $C_{R,D}\equiv m\lambda^2_{R,D}/\hbar^2$,
and the tilting angle $\theta$ enters the quantities ${\cal V}$,
${\cal Z}$, and ${\cal S}$ defined in the Appendix.
Tilting effects might arise because of the
$1 - |g^*| m^* {\cal S}/2$ denominator in the above equation, but
sizeable effects on $\Delta E_{CR}$ should only be expected for
materials such that  $|g^*| m^*/2$ is large. 
This is not the case for GaAs, but it is, e.g., for
InAs and InSb, which have $|g^*| m^*/2=0.169$ and 0.355, respectively.
For the latter case the dependence of $\Delta E_{CR}$ with 
the in-of-well field $B_x$, with a fixed $B_z$, is shown in Fig.\ 3. 
Notice that $\Delta E_{CR}$ 
is sharply increased when $B_x$ exceeds 
a given value (1T for the parameters in Fig.\ 3), which is proving 
the strong enhancement of SO effects introduced by the horizontal component 
of the tilted field configuration. 
Figure 3 also shows
the comparison with the exact diagonalization data (symbols), indicating that 
the analytical formula, Eq.\ (\ref{eq56}), is accurate up to rather large 
tilting angles and for varying relative weights of Rashba and Dresselhaus terms. 
As a matter of fact, this analytical result does not depend on $B_z$
although, for the sake of comparison with the exact diagonalization, we have used 
$B_z=1$ T in Fig.\ 3. The evolution with $B_x$ is not always monotonous, 
especially
for $C_R>C_D$ where we find an initial decrease of
$\Delta E_{CR}$ with increasing $B_x$, vanishing at $B_x\sim 0.8$ T, 
and eventually increasing again.

The tilting also affects the spin splitting of the Landau 
levels 
\begin{equation}
{\Delta E_n\over\omega_c}=
\frac{|g^*| m^*}{2} {\cal S} +
2(2 n + 1)\left[ (y_R~{\cal V} +y_D~{\cal Z})
\frac{1}{1 + |g^*| m^* {\cal S}/2} -
(y_R~{\cal Z}+y_D~{\cal V})
\frac{1}{1 - |g^*| m^* {\cal S}/2}
\right] \;\; , 
\label{eq57}
\end{equation}
which generalizes Eq. (\ref{eq51})  for $\theta\ne0$. 
As we have commented before, in a recent experiment
where spin precession frequencies in a InGaAs quantum well
 have been measured 
using electrically detected electron-spin resonances,\cite{Sih04}
a strong dependence of the effective gyromagnetic factor $g^{eff}$
on the applied tilted $B$  has been found. In particular, at
$\theta=45^o$  $g^{eff}$ exhibits oscillations with $B$ which indicate
its sensitivity to the Landau level filling, and a coupling
between spin and orbital eigenstates
which is explicitly present
in the spin-orbit term of Eq.(\ref{eq57}). The effective
g-factor
that can be extracted from this equation at $\theta=45^o$, 
by taking the ratio
$2\Delta E_n/(m^* {\cal S}\omega_c)$,
has the structure 
\begin{equation}
\vert g^{eff}(B,n)\vert= \vert g^*_0\vert + \left(n+\frac{1}{2}\right)
\left[c_1 B + {c_2\over B}\right] \;\;\;\; ,
\label{eq58}
\end{equation}
where the parametrization $g^*= g^*_0 -c_1(n+\frac{1}{2})B$ of
Refs. \onlinecite{Dob88,Sih04} has been introduced in Eq. (\ref{eq57}),
and the $c_2$ term is the SO contribution.
For the smaller $B$ values in the experiment,
and for reasonable values of  $m\lambda^2_{R,D}/\hbar^2$,
of the order of 1-10$\mu$eV, the SO contribution is 
important enough and should not be
neglected; under these circumstances, time-resolved Faraday
rotation spectroscopy could be sensible to Rashba and/or Dresselhaus
spin-orbit effects.

\section{Summary}

We have discussed the appearance of spin-orbit effects in
magnetoresistivity and inelastic light scattering experiments on
quantum wells. In particular, we have addressed SO effects on
the splitting of the cyclotron resonance, on the sp Landau level
spectrum, and  on spin-density excitations.
Our discussion has been based on the use of an analytical
solution of the quantum well Hamiltonian valid up to second
order in the SO coupling constants. The accuracy of this
solution has been assessed comparing it with exact numerical
diagonalizations.

We have carried out semi-quantitative comparisons with available
experimental data, with the twofold aim of extracting the value
of the SO coupling constants and of indicating possible manifestations
of the SO interactions. We have also pointed out that tilting the
-usually- perpendicularly applied magnetic field might enhance
spin-orbit effects, making them easier to detect.

\appendix*
\section{}

In this Appendix  we generalize some of the expressions
derived in Sec. II to the  case in which $B$ has a
in-of-well component, e.g., $B=(B_x,0,B_z)$.
The Zeeman term then
becomes $\frac{1}{2} g^* \mu_B {\bf B}\cdot$\mbox{\boldmath $\sigma$}=
$-\frac{1}{2} \omega^z_L (\sigma_x \tan\theta + \sigma_z)$,
where we have introduced  the zenithal angle $\theta$,
 $\tan\theta=B_x/B_z$, and the `$z$-Larmor' frequency
$\omega^z_L=|g^* \mu_B B_z|$, with  $\omega^z_L/\omega_c=|g^*| m^*/2$.
The Schr\"odinger Eq. (\ref{eq5}) then becomes
\begin{equation}
\left[ \begin{array}{cc}
\frac{1}{2}(a^+a^- + a^-a^+) - \omega^z_L/(2\omega_c)-\varepsilon
\;\;\;
& i\tilde{\lambda}_R a^- + \tilde{\lambda}_D a^+ 
-[\omega^z_L/(2\omega_c)]\tan\theta  \\
-i\tilde{\lambda}_R a^+   +\tilde{\lambda}_D a^- 
-[\omega^z_L/(2\omega_c)]\tan\theta  \;\;\; &
\frac{1}{2}(a^+a^- + a^-a^+) + \omega^z_L/(2\omega_c)-\varepsilon
 \end{array}\right]\left(\begin{array}{c}\phi_1\\\phi_2\end{array} \right)=0
\;\;\;\; .
\label{eqa1}
\end{equation}
The calculation proceeds as before, Eq. (\ref{eq6}) becoming
\begin{eqnarray}
(n+\alpha-\varepsilon)~b_n -{\alpha-\beta\over2}\tan\theta~a_n
-i\tilde{\lambda}_R\sqrt{n}~a_{n-1}+\tilde{\lambda}_D\sqrt{n+1}~a_{n+1}=0
\nonumber\\
(n+\beta-\varepsilon)~a_n -{\alpha-\beta\over2}\tan\theta~b_n
+i\tilde{\lambda}_R\sqrt{n+1}~b_{n+1}+\tilde{\lambda}_D\sqrt{n}~b_{n-1}=0  
\;\;\;\; ,
\label{eqa2}
\end{eqnarray}  
where $\alpha=(1+\omega^z_L/\omega_c)/2$ and $\beta=(1-\omega^z_L/\omega_c)/2$.

The sp spectrum Eq. (\ref{eq23}) becomes
\begin{eqnarray}
E_n^d=\left(n+{1\over2}\right)\omega_c+{\omega^z_L\over2}~{\cal S}
 + 2n\left[{\cal U}~(\lambda^2_{R}+ \lambda^2_{D})+
(\lambda^2_{R}~{\cal V}+\lambda^2_{D}~{\cal Z}){\omega_c\over
\omega_c+\omega^z_L~{\cal S}}\right]
\nonumber\\
- 2(n+1)\left[{\cal U}~(\lambda^2_{R}+
\lambda^2_{D})+
(\lambda^2_{R}~{\cal Z}+\lambda^2_{D}~{\cal V}){\omega_c\over
\omega_c-\omega^z_L~{\cal S}}\right]
\nonumber\\
E_n^u=\left(n+{1\over2}\right)\omega_c-{\omega^z_L\over2}~{\cal S} + 
2n\left[
{\cal U}~(\lambda^2_{R}+ \lambda^2_{D}) +
(\lambda^2_{R}~{\cal Z}+\lambda^2_{D}~{\cal V}){\omega_c\over
\omega_c-\omega^z_L~{\cal S}} 
\right]
\nonumber\\
- 2(n+1)\left[ {\cal U}~(\lambda^2_{R}+ \lambda^2_{D}) +
(\lambda^2_{R}~{\cal V}+\lambda^2_{D}~{\cal Z}){\omega_c\over
\omega_c+\omega^z_L~{\cal S}}
\right]  \;\;\;\; ,
\label{eqa3}
\end{eqnarray}
where we have defined ${\cal S}=1/\cos\theta$, ${\cal U}=\sin^2\theta/4$,
${\cal V}=(1+ \cos\theta)^2/4$, and ${\cal Z}=(1- \cos\theta)^2/4$.
When $\theta=0,\,{\cal U}= {\cal Z}=0,\, {\cal V}=1$, and
Eq. (\ref{eqa3}) reduces to Eq. (\ref{eq23}).

\section*{ACKNOWLEDGMENTS} This work has been performed 
under grants FIS2005-01414 and FIS2005-02796 from
DGI (Spain) and 2005SGR00343 from Generalitat de Catalunya.
E. L. has been suported by DGU (Spain), grant SAB2004-0091.

\pagebreak

\begin{figure}[t] 
\centerline{\includegraphics[width=14cm,clip]{fig2.eps}}
\caption{
Lower energy levels for a GaAs quantum well as a function of the
Dresselhaus intensity $y_D = \lambda_{D}^2/\omega_c$ 
 for a fixed  Rashba intensity $y_R =\lambda_{R}^2/\omega_c=0.01$.
  Solid
lines are the analytical result, Eq.\ (\ref{eq23}), while
symbols correspond to the exact diagonalization, Eq.\ (\ref{exd}).
} 
\label{fig2}
\end{figure}


\begin{figure}[t] 
\centerline{\includegraphics[width=14cm,clip]{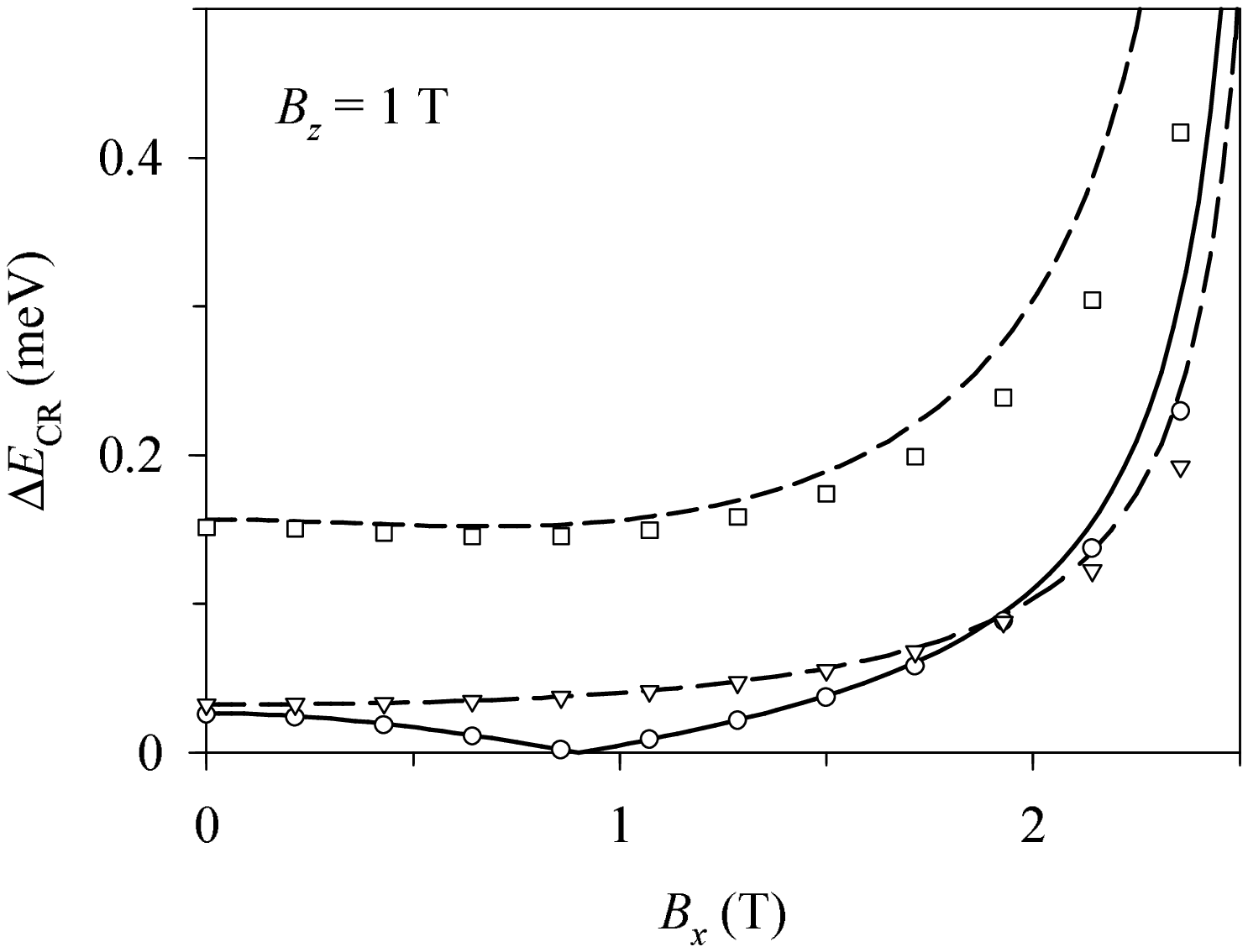}}
\caption{
Splitting of the cyclotron resonance for an InSb quantum well ($|g^*|m^*/2=0.355$) 
as a function of the in-of-well field $B_x$ when $B_z=1$ T.
Lines are the result from the analytical formula, Eq.\ (\ref{eq56}), while symbols correspond
to the exact diagonalization of Eq.\ (\ref{eqa1}).
Defining $C_{R,D}=m\lambda_{R,D}^2/\hbar^2$ the shown results are for:
$C_R=30$ $\mu$eV and $C_D=10$ $\mu$eV, solid line and circles; 
$C_R=10$ $\mu$eV and $C_D=10$ $\mu$eV, long-dashed line and triangles;
$C_R=10$ $\mu$eV and $C_D=30$ $\mu$eV, short-dashed line and squares.
} 
\label{fig3}
\end{figure}


\begin{thebibliography}{99}

\bibitem{Can99} C-M. Hu, J. Nitta, T. Akazaki, H. Takayanaga, J. Osaka,
P. Pfeffer, and W. Zawadzki, Phys. Rev. B {\bf60}, 7736 (1999).

\bibitem{Ric99} D. Richards and B. Jusserand, Phys. Rev. B {\bf59}, R2506 (1999).

\bibitem{And99} E. de Andrada e Silva, Phys. Rev. B {\bf60}, 8859 (1999).

\bibitem{Vos00} A. Voskoboynikov, S.S. Liu, C.P. Lee, and O. Tretyak, 
J. App. Phys. {\bf87}, 1 (2000).

\bibitem{Mal00} A.G. Mal'shukov and K.A. Chao, Phys. Rev. B {\bf61}, R2413 (2000).

\bibitem{Rac97} P.N. Racec, T. Stoica, C. Popescu, M. Lepsa, and T.G. van de Roer, 
Phys. Rev. B {\bf56}, 3595 (1997).

\bibitem{Vos01} O. Voskoboynikov, C.P. Lee, and O. Tretyak, Phys.\ Rev.\ B {\bf 63},
165306 (2001).

\bibitem{Fol01} J. A. Folk, S.R. Patel, K.M. Birnbaum, C.M. Marcus, C.I. Duru\"oz,
 and J.S. Harris, Jr, Phys.\ Rev.\ Lett.\ {\bf 86}, 2102 (2001).

\bibitem{Hal01} B.I. Halperin, A. Stern, Y. Oreg, J.N.H.J. Cremers, J.A.
Folk, and C.M. Marcus, Phys.\ Rev.\ Lett.\ {\bf 86}, 2106 (2001).

\bibitem{Ale01} I.L. Aleiner and V. I. Fal'ko, Phys.\ Rev.\ Lett.\ {\bf
87}, 256801 (2001).

\bibitem{Val02} M. Val\'{\i}n-Rodr\'{\i}guez, A. Puente, Ll.\ Serra, and 
E. Lipparini, Phys. Rev. B {\bf 66}, 165302 
(2002).

\bibitem{Val202} M. Val\'{\i}n-Rodr\'{\i}guez, A. Puente, and Ll. Serra, 
Phys. Rev. B {\bf 66}, 045317 (2002).

\bibitem{Val302} M. Val\'{\i}n-Rodr\'{\i}guez, A. Puente, Ll.\ Serra, and 
E. Lipparini, Phys. Rev. B {\bf 66}, 235322 
(2002).

\bibitem{Sch03} J. Schliemann, J.C. Egues, and D. Loss, Phys. Rev. B {\bf 67},
085302 (2003).

\bibitem{Kon05} J. K\"onemann, R.J. Haug, D.K. Maude, V.I. Fal'ko, and 
B.L. Altshuler, Phys. Rev. Lett. {\bf 94}, 
226404 (2005).

\bibitem{Cal05} M. Califano, T. Chakraborty, P. Pietilainen,
Phys. Rev. Lett. {\bf 94}, 246801 (2005).

\bibitem{Dre55} G. Dresselhaus, Phys. Rev. {\bf 100}, 580 (1955).

\bibitem{Ras84} Yu. A. Bychkov and E.I. Rashba, J. Phys. C {\bf 17}, 6039 (1984).

\bibitem{Pik95} F.G. Pikus and G.E. Pikus, Phys. Rev. 
B {\bf 51}, 16928 (1995).

\bibitem{Man01} M. Manger, E. Batke, R. Hey, K.J. Friedland, K. K\"ohler,
and P. Ganser,  Phys. Rev. B{ \bf 63}, 121203(R) (2001).

\bibitem{Ton04} P. Tonello and E. Lipparini, Phys. Rev. B{ \bf 70}, 081201(R) (2004).

\bibitem{Mal06} F. Malet, E. Lipparini, M. Barranco, and M. Pi, Phys. Rev. B
{\bf 73}, 125302 (2006).

\bibitem{Ste82}
D. Stein, K. v. Klitzing, and G. Weimann, Phys. Rev. Lett. {\bf 51},
130 (1982).

\bibitem{Dob88} 
M. Dobers, K. v. Klitzing, and G. Weimann, Phys. Rev. B {\bf 38}, 5453 (1988).

\bibitem{Dav97} H.D.M. Davies, J.C. Harris, J.F. Ryan, and A.J. Turberfield, 
Phys. Rev. Lett. {\bf 78}, 4095 (1997).

\bibitem{Kan00} M. Kang, A. Pinczuk, B.S. Dennis, M.A. Eriksson, 
L.N. Pfeiffer, and K.W. West, Phys. Rev. Lett. {\bf 84}, 546 
(2000).

\bibitem{Eri99} M.A. Eriksson, A. Pinczuk, B.S. Dennis, S.H. Simon, 
L.N. Pfeiffer and K.W. West, Phys. Rev. Lett. {\bf 82 }, 2163 (1999).

\bibitem{Sal01} G. Salis, D.D. Awschalom, Y. Ohno, and H. Ohno, 
Phys. Rev. B {\bf 64}, 195304 (2001).

\bibitem{Sih04} V. Sih, W.H. Lau, R.C. Myers, A.C. Gossard, M.E.
Flatt\'e, and D.D. Awschalom, Phys. Rev. B {\bf 70}, 161313(R) (2004).

\bibitem{Ras60} E.I. Rashba, Fiz. Tverd. Tela (Leningrad) {\bf 2}, 1224 (1960)
[Sov. Phys. Solid State {\bf 2}, 1109 (1960)].

\bibitem{Das90} B. Das, S. Datta and R. Reifenberger, Phys. Rev. B {\bf 41}, 8278
(1990).

\bibitem{Fal93} V.I. Falko, Phys. Rev. B {\bf 46}, R4320 (1992).

\bibitem{Val06} M. Val\'{\i}n-Rodr\'{\i}guez and R. G. Nazmitdinov,
cond-mat/0512231.

\bibitem{Kal84} C. Kallin and B.I. Halperin, Phys. Rev. B {\bf 30}, 5655 (1984).

\bibitem{Ser99} 
Ll.\ Serra, M. Barranco, A. Emperador, M. Pi, E. Lipparini,
Phys.\ Rev.\ B {\bf 59}, 15290 (1999).

\bibitem{Physrep} E. Lipparini and S. Stringari, Phys. Rep.
{\bf 175}, 103 (1989).

\bibitem{Lip04} E. Lipparini, 
{\it Modern Many Particle Physics-Atomic Gases, Quantum Dots
and Quantum Fluids} (World Scientific, Singapore 2003).

\bibitem{Mal86} F. Malcher, G. Lommer, and U. R\"ossler, Superlattices
and Microstructures {\bf 2}, 267 (1986);
G. Lommer, F. Malcher, and U. R\"ossler, {\it ibid.} {\bf 2}, 273
(1986).

\bibitem{Lip99} E. Lipparini, M. Barranco, A. Emperador, M. Pi, and
Ll. Serra, Phys. Rev. B {\bf 60}, 8734 (1999).

\end{thebibliography}
\end{document}